\documentstyle[prd,preprint,floats,epsfig,aps]{revtex}
\tightenlines
\def\dif{{\rm d}}
\def\rarrow{\rightarrow}
\def\lrarrow{\leftrightarrow}
\def\dnot{\! \not{\! \partial}}
\def\div{\vec \nabla \! \cdot \!}
\def\grad{\vec \nabla}
\def\LAG{{\cal L}}
\def\hc{{\rm h.c.}}
\def\sumli{\sum\limits}
\def\ie{\mbox{\em i.e.}}
\def\etal{\mbox{\em et.\ al.}}
\def\etc{{\em etc.}}
\def\ibid{{\em ibid}}

\def\I{\'\i{}}

\def\nnn{neutrino}
\def\nnns{neutrinos}
\def\ng{Nambu-Goldstone}
\def\gs{Goldstone}
\def\nue{{\nu_e}}
\def\numu{{\nu_\mu}}
\def\nutau{{\nu_\tau}}
\def\barnue{{\bar\nu_e}}
\def\barnumu{{\bar\nu_\mu}}
\def\barnutau{{\bar\nu_\tau}}
\def\U1{{\rm U(1)}}

\def\evs{\,{\rm eV}}
\def\kev{\,{\rm KeV}}

\def\mev{\,{\rm MeV}} 
\def\tev{\, {\rm TeV}}
\def\cms{\,{\rm cm}}
\def\sec{\,{\rm s}}
\def\any{Ann. N. Y. Acad. Sci.\ }
\def\apj{Astrophys. J.\ }
\def\arnps{Annu. Rev. Nucl. Sci.\ }

\def\npb{Nucl. Phys. B\ }
\def\nci{Nuovo Cimento\ }
\def\plb{Phys. Lett. B\ }

\def\prl{Phys. Rev. Lett.\ }
\def\prd{Phys. Rev. D\ } 
\def\rmp{Rev. Mod. Phys.\ }
\def\sjnp{Sov. J. Nucl. Phys.\ }

\def\yad{Yad. Fiz.\ }
\def\ga{\gamma}
\def\ffi{\phi}
\def\lll{\Lambda}
\def\vl{V_\lll}
\def\nul{\nu_L}

\def\le{{L_e }}
\def\lmu{{L_\mu }}
\def\ltau{{L_\tau }}
\def\al{A_\lll}
\def\alv{\vec \al}
\def\aev{\vec A_e}
\def\amuv{\vec A_\mu}
\def\atauv{\vec A_\tau}
\def\jlv{\vec J_\lll}
\def\dmjm{\partial_\mu J^\mu}
\def\jle{\vec j_e}

\def\jltau{\vec j_\tau}
\def\barnu{\bar \nu}
\def\GEE{G_{ee}}
\def\GEM{G_{e\mu}}
\def\GET{G_{e\tau}}

\def\GMT{G_{\mu\tau}}
\def\GTT{G_{\tau\tau}}
\def\ergss{{\rm ergs/sec}}
\def\ergssmev{{\rm ergs/Mev/sec}}
\def\df51{\delta F_{51}}
\def\m31{\tilde M_{31}}
\def\Enu{E_{10}}
\def\Dm2{\Delta m^2}
\def\dm2ct{\Delta m^2 \, \cos 2\theta}
\def\M2mu{M^2_{\mu}}
\def\nenu{\barnue \lrarrow \barnumu}
\begin{document}
\draft
\preprint{CFNUL/97-01}
\title{Neutrino Oscillations: a source of Goldstone fields}
\author{Lu\I s Bento}
\address{Centro de F\I sica Nuclear, Universidade de Lisboa,\\ 
Av. Prof. Gama Pinto 2, 1699 Lisboa - {\sl codex}, Portugal\\
{\rm e-mail: LBento@fc.ul.pt} }
\date{31 August 1997, to appear in Physical Review D 1 (1998).}
\maketitle
\begin{abstract}
 It is proved that true Goldstone bosons develop coherent fields whenever
 the associated charges of the matter particles are not conserved in
 a macroscopic scale.
 The sources of the Goldstone fields are the time rates of quantum number
 violation.
 The case of \nnn\ flavour oscillations is studied with application to
 Supernovae.
 It is shown that if the Lepton numbers break at the Fermi scale, the
 $\nu$ potentials and oscillation patterns change in the periods of
 largest $\nu$ fluxes.
 In this way, $\barnue \lrarrow \barnumu$ oscillations may occur in the
 first instants of $\nu$ emission.
\end{abstract}


\section{Introduction}

In theories with spontaneous breaking of global symmetries the \gs\
theorem asserts the existence of massless particles - the \gs\
bosons - one per each non-zero quantum number of the vacuum.
In principle such massless scalar bosons could mediate long range forces
and lead to observable coherence effects just like the gravitational
force.
However, the very structure imposed by exact global symmetries at the
Lagrangian level implies that these goldstons have nothing but
derivative couplings with matter fields.
It was shown that if the symmetry constraints are relaxed either through
anomalies or with soft symmetry breaking terms then, scalar couplings
of pseudo-goldstons with fermions naturally arise.
Consequently, macroscopic forces are developed by matter aggregates but
one price is paid: the intermediate bosons get non-zero masses which
means that the range of those forces is finite.
In this paper it is shown that under special conditions a true \ng\
boson may develop a macroscopic field, the source being the rate of
non-conservation of the matter quantum number associated with that field.
That may be significant in astrophysical compact objects and the case of
supernovae is analysed.

To be definite I consider an abelian global symmetry $\rm{ U(1)_\lll}$ 
 where $\lll$ denotes generically any linear combination of flavour
lepton numbers such as $\le -\lmu$ or the total lepton number $L$.
After symmetry breaking non-invariant terms are generated in 
masses and/or couplings but the symmetry is still realized non-linearly
and strongly constrains the couplings to the \gs\ boson $\ffi$.
 Under $\rm{ U(1)_\lll}$, the matter fields and $\ffi$ transform as
	 \begin{mathletters}
 \label{trfs}
 	\begin{equation}
\psi^a \rarrow e^{-i\, \alpha \, \lll_a} \psi^a
	\, ,
	\end{equation}					\label{tpsi}
 	\begin{equation}
\ffi \rarrow \ffi + \alpha\, \vl
	\, ,						\label{tfi}
	\end{equation}
	\end{mathletters}
 respectively, where $\vl$ is the scale of symmetry breaking.

Consider for instance the lepton masses.
Let $\nul^a$ denote all left-handed neutrino fields including the standard
weak interacting partners $\nul^e,\nul^\mu,\nul^\tau$ and any extra neutral
singlets.
The mass terms generated either radiatively or at tree-level are
necessarily accompanied by couplings to $\ffi$ completely
determined by the lepton charges $\lll_a$:
	\begin{eqnarray}
\LAG_{\rm m} = - {m_{ab}\, {\nul^a}^T \kern1pt C\, \nul^b
\; e^{i\left( {\lll_a+\lll_b} \right)\ffi /\vl } }
+ \hc -  m_\ell \, {\bar \ell}\, \ell
	\, ,						\label{Lm}
	\end{eqnarray}
where a sum over $a,b$ and $\ell=e, \mu, \tau$ is assumed.
Moreover, in any other term of the Lagrangian each field $\psi^a$ always
comes with a factor $\exp ({i\, \lll_a \ffi /\vl })$.
Consequently, a spatially uniform $\ffi$ would be absorbed in the mass
eigenstate fields leaving no couplings with $\ffi$.
Matter only cares about the gradient of $\ffi$ and that, as shown in
\cite{gelm83}, is better expressed in what I term as weak-unitary basis:
	\begin{eqnarray}
\psi^a \,( {\rm old} ) = e^{-i\, \lll_a \ffi /\vl }\, \psi^a
\, ( {\rm uni} )
	\, .						\label{psi}
	\end{eqnarray}
The new fields preserve the original quantum number assignments but are
invariant under the global symmetry transformations (\ref{trfs}).
They reduce now to a translational invariance
$\ffi \rarrow \ffi + \beta$ and therefore, the goldston interactions
can only depend on the $\ffi$ derivatives.

In the unitary basis the effective Lagrangian writes as
	\begin{eqnarray}
\LAG = & \sum {\bar \psi}i \dnot \psi -
\sum\limits_{e,\mu,\tau} m_\ell \, {\bar \ell}\, \ell 
- \sum\limits_{a,b} { {\nul^a}^T \, C\, m_{ab}\, \nul^b \, } + \hc &
  	\nonumber\\
 & \mbox{ } +{1 \over 2}\, \partial_\mu \ffi \, \partial^\mu \ffi + 
J_\lll^\mu \, \partial_\mu \ffi /\vl + \cdots 
	\, 	&					\label{Luni}
	\end{eqnarray}
All of the goldston interactions are condensated in the current
	\begin{eqnarray}
J_\lll^\mu = \sumli_{e,\mu,\tau} {\bar \ell}\ga^\mu \lll_\ell \,\ell
 + \sum\limits_{a} {\bar \nul^a}\ga^\mu \lll_a \, \nul^a + \cdots
	\, ,						\label{jl}
	\end{eqnarray}
which includes the matter current derived from the kinetic Lagrangian of
$\psi \, ( {\rm old} )$ but also in general model dependent higher
order corrections.
The $\ffi$ equation of motion reads
	\begin{eqnarray}
\partial_\mu \partial^\mu \, \ffi = - \dmjm_\lll /\vl
	\, 						\label{ddfi}
	\end{eqnarray}
and clearly the source term is not some sort of matter density.

The divergences of $\bar \psi \gamma^\mu \psi$ and
$\bar \psi \gamma^\mu \gamma_5 \psi$ only yield pseudo-scalar
$\bar \psi \gamma_5 \psi$ charges that do not generate long range '$1/r$'
 fields, but rather spin dependent '$1/r^3$' potentials \cite{chik81}.
In the intent of escaping the derivative couplings law, models were
made with explicit symmetry breaking terms.
These may be anomaly terms \cite{wilc82}, as in the axion case
\cite{pecc77}, or soft symmetry breaking terms \cite{hill88}.
In this way scalar couplings with the goldston field are produced but
at the cost of getting a mass for the so-called pseudo Nambu-Goldstone
boson which ultimately means a finite range for the long range forces.

Here, I will explore the fact that the source term is different from zero
whenever the $\lll$ number of the matter particles is not conserved.
To be accurate, there is a No\"ether current associated to the symmetry
$\ffi \rarrow \ffi + \alpha$ namely $J^\mu_\lll + \vl \partial^\mu \ffi$,
that is exactly {\em conserved},
 but the non-standard interactions do not conserve the fermions $\lll$
 number as a result of the very spontaneous symmetry breaking.
 
Non-conservation of $J^\mu_\lll$ may occur in macroscopic proportions
in a stationary basis in stars, or in Supernovae for a short period of time.
For instance, the total lepton number is violated at a constant rate in
the process of neutrino emission of a star if $\nu_e$ has a Majorana mass:
the non-conservation takes place in the neutrino production reaction where
a fraction of the neutrino wave function is an anti-neutrino $\barnu_e$.
This fraction is of course suppressed by the ratio $m_\nu^2 /E_\nu^2$.
That is not the case if a significant portion of the $\nu_e$ spectrum is
converted into $\barnumu$ anti-neutrinos under appropriate magnetic field
 and transition magnetic moments.
Another example is a \gs\ boson associated with the breaking of
$\le - \lmu$.
If there is resonant conversion of $\nu_e$s into $\numu$s then the
goldston source has macroscopic dimensions and a long-range
scalar field is generated.

\section{Goldstone fields}

The solution of the $\ffi$ equation of motion is found using standard
Green function methods.
In the source term, $\dmjm_\lll = \dot \rho _\lll ^{\rm cr}$ represents
the rate of {\em creation} of $\lll$-charge per unity of time and volume.
In a stationary regime where this rate is constant or varies
slowly enough one gets a Coulombian field 
	\begin{eqnarray}
\ffi ( t, \vec r ) = {{-1} \over {4\pi }} \int {d^3x \,
{\langle \dmjm_\lll ( t,\vec x ) \rangle
\over {\vl \left| {\vec r-\vec x} \right|}}}
	\; ,						\label{field}
	\end{eqnarray}
where the bracket means expectation value of the operator $J^\mu_\lll$
over the appropriate particle state density.
The Lagrangian of Eq.\ (\ref{Luni}) tells however that the fermions couple  
to the gradient $\al^\mu = \partial^\mu \ffi /\vl$.
In the stationary limit $\al^0$ is negligible and
the vector $\alv =-\grad \ffi /\vl$ obeys a Gauss law:
	\begin{eqnarray} 
\div \alv = - \langle \dot \rho_\lll + \div \jlv \rangle /\vl^2
	\, \, .						\label{diva}
	\end{eqnarray}
It implies that the flux of $\alv$ over a closed surface is proportional
to the time rate $\dot Q_\lll^{\rm cr}$ of {\em creation} of $\lll$-charge
inside the surface.
In turn $\dot Q_\lll^{cr}$ is equal to the difference between the flux of
$\langle \jlv \rangle$ and the flux of the current $\jlv^0$
that would exist if there was no $\lll$ violation inside the surface.
Hence,
	\begin{eqnarray}
\alv = - \left( \langle \jlv \rangle -  \jlv^0 \right) /\vl^2
	\, .						\label{alv1}
	\end{eqnarray}
The value of $\jlv^0$ is determined for a given $\dot \rho_\lll$
and obeys the identity $\dot \rho_\lll  + \div  \jlv^0 =0$.
One reaches the same solution by noting that the No\"ether current
$J^\mu_\lll + \vl \partial^\mu\ffi$ is conserved.

A typical situation of potential interest is the flavour violation
driven by \nnn\ oscillations in their way out of a star.
The quantum numbers are the partial lepton numbers $L_\ell=\le,\lmu,\ltau$
or linear combinations of them and the bare currents are
	\begin{eqnarray}
j_\ell^\mu = {\bar \ell}\ga^\mu \, \ell +
{\bar \nul^\ell} \ga^\mu \, \nul^\ell +
\sum\limits_{a>\tau} {\bar \nul^a} \ga^\mu (L_\ell)_a \, \nul^a
	\; ,						\label{jll}
	\end{eqnarray}
where $\nul^a ,$ $a>\tau$, denote any non-standard neutrinos.
To be specific suppose that $\ltau$ is spontaneously broken.
The \gs\ boson $\ffi_\tau$ associated with it couples to a current
$J_\lll^\mu$ that is essentially $j_\tau^\mu$ but contains in general
higher order terms arising from radiative corrections.
Their existence implies that even if $\ltau$ is conserved in a
certain process, the very reaction rate \ie, of particle production or
absorption, is enough to yield a goldston field.
However those terms are of higher order and I will limit myself
to the bare tree-level currents.

For definiteness I assume that all non-standard \nnns\ are extra-heavy
singlet \nnns\ that cannot be produced at energies of interest
and in addition, no significant \nnn\ $\lrarrow$ anti-\nnn\ transitions
occur either because the light \nnn\ masses are 
too small or because transition magnetic moments do not come into play.

Suppose that in the way out of a star $\nu_e$s convert into $\nutau$s
and the flux of converted \nnns\ is $\vec j (\nu_e \rarrow \nutau)$.
Then, a macroscopic goldston field $\ffi_\tau$ exists with non-zero gradient
outside the sphere where the conversion takes place.
In the Supernovae case one also gets a flux $\vec j (\nutau \rarrow \nu_e)$.
Eq. (\ref{alv1}) gives $\atauv = -  \delta \jltau /{V_\tau^{2}}$ with
	\begin{eqnarray}
\delta \jltau = \vec j (\nu_e \rarrow \nutau)- \vec j (\nutau \rarrow \nu_e)
	\, .						\label{atauv}
	\end{eqnarray}
If the conversion involves anti-neutrinos one has instead
$\delta \jltau = -  j (\barnu_e \rarrow \barnu_\tau) +
\vec j (\barnu_\tau \rarrow \bar\nu_e)$,
because the antineutrinos have negative Lepton numbers.

The field $\ffi_\tau$ yields a potential energy for the $\tau$ \nnns.
Generalising now to any quantum number $L_\ell$ and goldston field
 $\ffi_\ell$, the interaction Lagrangian
$\LAG_{\rm int} = J_{\ell\,\mu} \, A_\ell^\mu$ in Eq.\ (\ref{Luni})
gives for the standard \nnns\ $\nu_\ell$,
	\begin{eqnarray}
\LAG = \bar \nul^\ell \ga_\mu \nul^\ell \, A_\ell^\mu,
 \quad\quad	 \ell=e, \mu, \tau		 	\, .
 	\end{eqnarray}
The potential energy of a neutrino with velocity $\vec v_\nu$ is therefore
	\begin{eqnarray}
V_{ \nu_\ell} = - A^0_\ell + \vec A_\ell \cdot \vec v_\nu
= - V_{\barnu_\ell} 
	\; .						\label{vnul}
	\end{eqnarray}
In the stationary limit $A^0_\ell =0$ and for radially moving neutrinos
$V$ is just the radial component $A_\ell^r$.
The interesting point is that such kind of potential is not universal in
flavour and in addition the sign of the potential is not solely determined
by the matter content of the medium where neutrinos propagate as it also
depends on the flavour structure of the \nnn\ fluxes.
This allows in principle other than the standard MSW type of oscillations.
The most interesting effect is probably the occurrence of
$\barnu_e \lrarrow \barnu_x$ in a situation where it would not
normally occur \ie , with a mass hierarchy
$m_{\nu_e} < m_{\numu}, \, m_{\nutau}$.
In the following I model such kind of phenomenon with application
to Supernovae \nnns .

If a partial Lepton number is spontaneously broken it is natural that
the same happens for all the three flavours.
Let $\sigma_i $ be the set of scalar fields,
singlets of SU(2)xU(1), with quantum numbers $\Lambda_i=(\le, \lmu, \ltau)_i$
 under the symmetry group $\U1_e \times \U1_\mu \times \U1_\tau$.
I assume that the vacuum expectation values $\langle \sigma_i \rangle$
break all the Lepton numbers producing 3 Goldstone bosons.
By expanding around the vacuum as
$\sigma _i=\exp \{ {-i\sumli_\Lambda \,{\xi ^\Lambda \Lambda _i}} \}
( {\langle \sigma_i \rangle +\cdots } )$,
where the dots stand for the massive degrees of freedom, one obtains for
 the Lagrangian of the \gs\ fields
	\begin{eqnarray}
\LAG={1 \over 2}V_{\Lambda M}^2\,
\partial _\mu \xi ^\Lambda \,\partial ^\mu \xi ^M +
J_\Lambda ^\mu \,\partial _\mu \xi ^\Lambda
	\, ,						\label{lxi}
	\end{eqnarray}
a sum over the quantum numbers $\Lambda, M$ is implicit.
The matrix
	\begin{eqnarray}
V_{\Lambda M}^2= 2\sum\limits_i {\Lambda _i}M_i
\left| {\langle \sigma_i \rangle } \right|^2
	\end{eqnarray}
is in general non-diagonal, the sole condition being that some of the fields
$\sigma _i$ have simultaneously more than one quantum number different
 from zero.
Such  mixing implies that the \gs\ bosons can mediate an action at a
distance between two distinct currents as the equations of motion show: 
	\begin{eqnarray}
\partial _\mu \partial ^\mu \xi ^\Lambda =- \sumli_M G_{\Lambda M}\, \dmjm_M
	\, ,						\label{ddxi}
	\end{eqnarray}
where $G_{\Lambda M}$ is the inverse matrix of $V_{\Lambda M}^2$.
Hence, in a stationary regime the vector field solutions are now
	\begin{eqnarray}  
\alv =- \sum\limits_M  G_{\Lambda M} 
\left( \langle {\vec J_M} \rangle -\vec J_M^0  \right)
	\, 						\label{alv2}
	\end{eqnarray}
instead of Eq.\ (\ref{alv1}).
The expression of the one-particle potential remains the same as in
Eq.\ (\ref{vnul}) because the interaction Lagrangian is the same as before namely, 
$J_\Lambda ^\mu \,A_{\Lambda \,\mu }=
J_\Lambda ^\mu \,\partial _\mu \xi ^\Lambda$.

 It is not the aim of this paper to construct specific models but as
 a proof of existence I present here a suitable adaptation of the abelian
 familon \cite{grin85} and seesaw Majoron \cite{chik81} models.
  Three neutral singlets $N_R^e, N_R^\mu, N_R^\tau\,$ join
 the standard lepton iso-doublets $L^e, L^\mu,$ $ L^\tau$, the charged
 singlets  $e_R, \mu_R, \tau_R$ and the Higgs doublet $\Phi$.
 The fields $N_R^\ell$ only have Yukawa interactions namely, flavour
 conserving Dirac couplings 
	\begin{eqnarray}
 \LAG_D = -
 {\left( {m_\ell \,\bar L^\ell \kern1pt \Phi\kern1pt \ell _R +
\mu_\ell\,\bar L^\ell\kern1pt \tilde \Phi\kern1pt N_R^\ell} \right)}/
\langle {\Phi ^0} \rangle + \hc
	\; ,
	\end{eqnarray}
 plus Majorana couplings with a set of scalar singlets
 $\sigma _{\ell m}=\sigma_{m\ell}$:
	\begin{eqnarray}
 \LAG_M = - {1 \over 2}
 \sumli_{\ell ,m} \left( {N_R^\ell}^T C\kern1pt N_R^m \right) \,
 M_{\ell m} \sigma _{\ell m}/\langle {\sigma _{\ell m}} \rangle
 	\; .
	\end{eqnarray}
  The invariance under $\U1_e \times \U1_\mu \times \U1_\tau$ determines
 the $(\le, \lmu, \ltau)$ quantum numbers of the scalar fields namely,
 $\sigma_{ee}$$(-2,0,0)$, $\sigma_{e\mu}$$(-1,-1,0)$,
 $\sigma_{\mu\tau}$$(0,-1,-1)$, \etc\
 The expectation values of these fields break all the three
 Lepton numbers which gives three \gs\ bosons.
 The matrix $G_{\ell m}$ that enters in the goldston equations of motion,
 Eq.\ (\ref{ddxi}), has full generality with respect to its flavour structure.
 For Majorana masses $M$ much higher than the Dirac masses $\mu_\ell$,
 the \nnns\ separate in heavy $N^\ell$ with mass matrix $M$ and very
 small mixing with the three light \nnns, $\approx \nul^\ell$,
 that acquire a Majorana mass matrix
 $m_{\ell m}=-\mu _\ell \kern1pt \mu _m\,M_{\ell m}^{-1}$.
 The magnitudes relevant for the subsequent application lie around
 $m \sim 10^{-2}\evs$, and $M =0.1-1\tev$, $\mu \sim 30 \kev$,
 very far from the original \cite{gell79} seesaw GUT scales but not excluded
 from first principles.
 
It is known that this type of singlet Majoron or familons easily escape
present astrophysical bounds on the couplings of Nambu-\gs\ bosons.
Here, they couple primarily to \nnns\ with pseudo-scalar couplings
$g \sim m_\nu /\vl$ which, for the specified range of \nnn\ masses,
 are far below $10^{-4}$ to play a role in Supernova collapse dynamics
 \cite{kolb82}, and even below the limit of $\sim 10^{-8.5}$ to
 produce Supernova cooling through singlet Majoron emission \cite{choi90}.
On the other hand, pseudo-scalar couplings to electrons, that could be
responsible for energy loss in stars \cite{raff97}, only arise through
 radiative corrections and are so further suppressed.

\section{Supernova neutrinos}

If the cause of the well known solar neutrino anomalies lies in \nnn\
flavour oscillations then one expects that the same phenomenum occurs
with Supernovae \nnns.
In the accepted understanding of Supernova type-II all neutrino species
including their anti-particles are emitted with similar importance after
a first electron \nnn\ burst.
As a working frame I adopt here the common prejudice that the \nnn\ mass
eigenstates are in first approximation - but not exactly - 
$\nu_1\approx \nu_e,\, \nu_2\approx \numu, \,\nu_3\approx \nutau$,
 with the following mass hierarchy: $m_1 < m_2 < m_3$.
In their way out from the star the electron \nnns\ have a potential
energy that exceeds by $V_W$ the equal potentials of $\numu$ and $\nutau$.
$V_W=\sqrt 2\, G_F \, n_e$ \cite{wolf78,kuo89} is positive and proportional
 to the electron density $n_e$ and therefore the \nnns\ propagate from the
region of emission where $V_W$ is much larger than $m_3$ (so I assume) to
the outer space where $V_W$ goes to zero.
It is then in principle possible to have  MSW resonant conversions from
$\nu_e$ into $\nutau$ and vice-versa \cite{beth86,walk87}.
 The anti-particle $\barnu_e$ on the contrary, receives a negative potential
 $-V_W$ and cannot be {\em resonantly} converted into the other flavours,
 given the assumed mass hierarchy. 

In a Supernova the $\nu_e$ flux is larger than the
$\nutau$ flux and consequently more $\nu_e$ transform into
$\nutau$ than the reverse so making a net $\le -\ltau$ violation.
As a result, the goldston fields
$\ffi_e, \,\ffi_\mu,\, \ffi_\tau$ develop as described above.
Applying Eq.\ (\ref{alv2}) with flux transfers
	\begin{eqnarray}
\delta \jltau = - \delta \jle =
\vec j (\nu_e \rarrow \nutau) - \vec j (\nutau \rarrow \nu_e)
	\, ,						\label{djtau}
	\end{eqnarray}
one derives the vector potentials
	\begin{eqnarray}
\vec A_\ell = (G_{e\ell} -G_{\ell\tau}) \, \delta \jltau \,,
\kern 50pt \ell =e,\mu,\tau   \;.			\label{alv3}
	\end{eqnarray}
They vanish of course inside the sphere where
$\nu_e \lrarrow \nutau$ takes place.
The magnitude of the constants $G_{\Lambda M}$ is directly given by
the inverse square of the Lepton symmetry breaking scale,
but the flavour structure is quite arbitrary and is not directly related
to the \nnn\ masses as they also depend on the Yukawa couplings.
The matrix $G_{\Lambda M}$ is only constrained to be symmetric and 
have positive eigenvalues as demands the goldston kinetic Lagrangian.

An interesting case is the following hierarchy:
$|\GEE|, |\GEM|, |\GET|$ $ << |\GMT|, \GTT$.
Then, the vector $\aev$ is much smaller in magnitude than $\amuv$
and $\atauv$.
 Far enough from the neutrinosphere and assuming spherical symmetry,
the $\nu$ potentials are well approximated by
	\begin{eqnarray}
& & V_{\nu_e} = V_W + A_e \,, \kern 40pt |A_e| << |V_\numu |
			\;, \nonumber\\
& & V_\numu = A_\mu \simeq - \GMT \, \delta j_\tau  	  \;, \label{vnu2}\\
& & V_\nutau = A_\tau \simeq - \GTT \, \delta j_\tau < 0   \;, \nonumber
	\end{eqnarray}
ignoring the universal neutral current and gravitational potentials.
The anti-\nnn\ potentials are the symmetric of these.

In the following the numbers for the \nnn\ fluxes are extrapolated from
the Supernovae \nnn\ luminosities obtained in \cite{mayl87} with a
$25 M_\odot \,C$ model and in the generic model of \cite{burr92}.
The energy fluxes follow this pattern:
$\numu, \nutau, \barnumu, \barnutau$ have all the same luminosity say
$L_\nutau$, and $\barnue$ gets rapidly close to the $\nue$ luminosity
 after the initial $\nu_e$ burst (instant $t=0$).
From then to the supernova explosion
($t \approx 0.45$ sec), the average 
value of $L_\nue$ is around $(4 - 8) \times 10^{52} \ergss$ and
$L_\nutau$ is about $2/3$ to $3/4$ of $L_\nue$.
After the explosion the spectrum is thermal and the luminosities
$L_\nutau \approx L_\barnue \approx L_\nue$ decrease along the time.
At $t= 1$ sec, $L_\nue = (4 - 10) \times 10^{51} \ergss$.
The fluxes are obtained from the number of emitted particles per second
$F=j(4\pi \kern 1pt r^2)$, and $F = L/\bar E$ by definition
of average energy $\bar E$.
I take $\bar E = 10 \mev$ for $\nue$ and $20 \mev$ for $\numu,
\nutau$.
If the resonance is adiabatic \ie, all $\nue$
transform into $\nutau$ and viceversa, the fluence transfer is
$\delta F_\tau = F_\nue -F_\nutau$.
Then, in unities of $10^{51} \ergssmev$, $\df51 \simeq 2 - 5$ 
before the explosion and $\df51 \simeq 0.2 - 0.5$ at 1 sec time.
In these unities,
	\begin{eqnarray}
V_{\numu}=
- 1.48 \; \GMT^F {{\delta F_{51}} \over {r_{10}^2}} \times 10^{-12}\, \evs
	\;,
	\end{eqnarray}
where $\GMT^F = \GMT /G_F$, $G_F=11,66 \tev^{-2}$ is the Fermi constant
and $r_{10} = r/10^{10}\cms $.

The standard charge current potential $V_W$ is proportional to the electron
density $n_e$ given by $Y_e \simeq 1/2$ times the nucleon
density $\simeq \rho /m_n$.
At large distances the mass density $\rho$ behaves as $\tilde M  /r^{3}$
with a constant $\tilde M = \m31 \times 10^{31} {\rm g}$, $\m31 =1 - 15$
depending on the star \cite{wils86}.
Then,
	\begin{eqnarray}
V_W=\sqrt 2\kern 1pt G_F \, n_e=
7.6 \,Y_e{{\tilde M_{31}} \over {r_{10}^3}}\times 10^{-13}\, \evs
	\; .
	\end{eqnarray}
$V_W$ decays necessarily faster than the goldston potentials which being
proportional to the $\nu$ fluxes inevitably overcome at large enough radii.
 That is a crucial point. 
The effective $M^2$ matrix \cite{kuo89} relevant for $\barnue-\barnumu$
oscillations is, up to a universal term, given by:
	\begin{eqnarray}
\left(     \begin{array}{cc}
M^2_e = 2E(V_\barnue -V_\barnumu) &\quad M^2_{e\mu} = \Dm2 \sin 2\theta	\\
				&					\\
M^2_{\mu e} = \Dm2 \sin 2\theta   &\quad \M2mu = \Dm2 \cos 2\theta	
\end{array}	\right)
	\; .
	\end{eqnarray}
An electron anti-neutrino is emitted with a negative potential
$V_\barnue =-V_W$ and a resonance would not be possible because $\Dm2 >0$,
$\theta < \pi/4$, by assumption.
But, if the constant $\GMT$ is negative, $\barnumu$ gets a negative
goldston potential $V_\barnumu = \GMT\, \delta j_\tau$ and the resonance
condition $M^2_e = M^2_\mu$ may be met.
In the Fig. 1 it corresponds to the interception of the
"e" and "$\mu$" curves.


\begin{figure}[h]
\vspace{15pt}
\centering
\epsfig{file=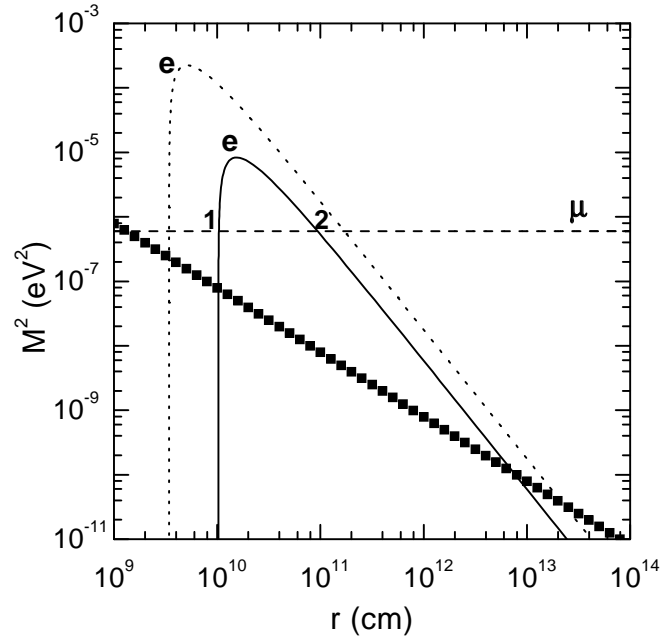,width=3.375in}
\vspace{10pt}
\caption{$2E(V_\barnue -V_\barnumu)$ is plotted in the solid and dotted
 curves with $\df51 =1$ and 3 respectively.
The dashed line stands for a particular value of $\M2mu=\Dm2 \cos 2\theta$
and the squares curve is the adiabatic boundary for $\M2mu \tan^2 2\theta$
in resonance 2.}
\end{figure}

The positive part of the 'potential' $M^2_e$ is plotted in Fig. 1 with
values $\GMT^F =-1$, $\df51=1$ or 3, $Y_e=1/2$, $\m31=4$, and a \nnn\ energy
 of $20 \mev$, the same for all curves.
Take $\M2mu = 6 \times 10^{-7} \evs^2$ for example.
 Its dashed line crosses the relative potential curve at two points: the
first, $r_1$, lies in the remarkably sharp transition from the $-V_W <0$
 dominated region to the goldston dominated one and very likely a
 non-adiabatic level crossing occurs there \ie, $\barnue$ continues $\barnue$
 and $\barnumu$ continues $\barnumu$.
 In the second point, $r_2$, the potential is smooth enough ($\propto r^{-2}$)
 for adiabatic transitions.
In general, the conversion probability is enhanced \cite{kuo89,mikh86}
 if $\M2mu\,\tan^2 2\theta$
is larger than the rate $f(r)=|E\,\dif V/V\dif r|$ at the resonance position.
This function is plotted with squares in the figure for $V \propto r^{-2}$.
One can see that small mixing angle adiabatic conversions are possible
for a wide range of \nnn\ masses.
Using the resonance condition (neglecting $V_W$) one gets a
lower bound on the mixing angle,
	\begin{eqnarray}
\M2mu \, \tan^4(2\theta) > 
 \frac{0.52 \, \Enu}{|\GMT^F \, \df51 |} \times 10^{-10} \evs^2
	\; ,						\label{m2mut4}
	\end{eqnarray}
while the maximum value of the $\nu$ potential sets an
upper limit to the square mass difference:
	\begin{eqnarray}
\M2mu < 1.66\, E_{10} \frac{|\GMT^F \, \df51 |^3}{(Y_e\, \m31)^2}
\times 10^{-5} \evs^2
	\; ,						\label{m2mu}
	\end{eqnarray}
where $\Enu=E/ 10\mev$.

 The requirement of non-adiabatic crossing in the first sharp
 resonance puts an upper limit on $\M2mu \tan^2 2\theta$.
 The slope of the potential on that point is equal (up to a numeric
 factor between 1 and 1/2) to the slope on the second resonance times the
 ratio $(r_2 /r_1 )^3$ and therefore that is the size of the window of
 interesting $\tan^2 2\theta$ values for a given $\M2mu$. 
 The net result is that for a wide range of parameters $\barnue$, $\barnumu$
 undergo one level crossing with maximum probability $P_c = 1$
 \cite{haxt86,park86} plus one adiabatic resonance.
 Their survival probabilities \cite{kuo89,park86} are therefore
 $\sin^2 \theta$ and not $\cos^2 \theta$ (mixing with $\nutau$ neglected).

 As far as $\nu_e$, $\numu$ is concerned, the potentials and $M^2_e$
are symmetric to the anti-neutrino case.
There is still one resonance position as in the standard model but now
in the abrupt part of the potential.
It results in a non-adiabatic crossing and
 the oscillation pattern is $\nu_e \rarrow \nutau$ and
$\nutau \rarrow \nu_e$, $\numu \rarrow \numu$ in contrast to the 'standard'
one \cite{beth86,walk87},
$\nutau \rarrow \nu_e \rarrow \numu$, $\numu \rarrow \nu_e$,
for the same mass mixing pattern.
This does not change the observation predictions because $\numu$, $\nutau$
have the same energy spectrum.
Finally, the late $\barnue \lrarrow \barnumu$ oscillations alter the net
flavour violation and modify the goldston fields out of the sphere
where they take place but that does not create new resonances.

This all picture may relate to solar \nnns\ in two different ways:
 1) $\nu_e \rarrow \nutau$
 in the Sun with $10^{-4} \evs^2 \geq \Dm2_{e\tau} > \Dm2_{e\mu}$.
Then, $\nu_e \rarrow \nutau$ ($\nutau \rarrow \nu_e$) triggers the
appearance of goldston fields in the supernova
(only out of the sphere where $\nu_e \rarrow \nutau$ takes place
in particular, if $\Dm2_{e\tau} \sim 10^{-5} \evs^2$, 
$r \gtrsim 2\times 10^{10}\cms$)
and eventually $\barnue \lrarrow \barnumu$ occurs.
2) $\Dm2_{e\tau}$ lies above $10^{-4} \evs^2$ and it is
$\nu_e \rarrow \numu$ that takes place in the Sun.
Then, resonant $\barnue \lrarrow \barnumu$ transitions occur in the supernova
with the kinematic parameters proper of solar \nnns.
 The solar non-adiabatic solution \cite{haxt86,park86,rose86} 
 with parameters $\M2mu \leq 10^{-5}$$ \evs^2$ and
$\M2mu \tan^2 2\theta \approx 4\times 10^{-8} \evs^2$ satisfies the desired
condition if the \nnn\ fluxes are high enough (compare the
dotted $\df51=3$ curve and the solid one, $\df51=1$).

This points to a distinctive feature of that phenomenon: the
strong dependence on the \nnn\ fluxes (third power in the bound
 of Eq.\ (\ref{m2mu})) should manifest by the detection of $\barnue$s in the
first moments of supernova
observation with a energy spectrum proper of $\barnumu$ ($25\mev$ instead
of $15\mev$) that is not maintained later when the fluxes drop down.
A crucial condition is of course that the scale of Lepton numbers
 symmetry breaking lies at or below the Fermi scale and that the goldston
potentials have the right sign.
The observation of the described correlation between oscillations and
 flux magnitudes would provide a measurement of the breaking scale.

 The analyses of the SN1987A events \cite{hira87} have either disfavoured the
 occurrence of $\nenu$ oscillations by comparing the best-fit energy spectrum
 with the current predictions for $\barnue$ and $\barnumu$ spectra
 \cite{jege96} or put a maximum limit of $35\%$ for the $\nenu$
 permutation probability \cite{smir94}.
 However, the oscillations driven by goldston
 fields, if they happen at all, only occur in the periods of largest fluxes.
 If they occur in the first 1/2 second or so of emission, the resultant hotter
 spectrum may manifest as a short artificial cooling time scale
 on the top of the real \nnn\ spectrum evolution.
 The fits to the SN1987A time evolution indeed exhibit two time scales,
 one of about $4.0\sec$ and another $\leq 1.0\sec$ \cite{burr90}.
 This seems consistent with an intrinsic two phase cooling process
 \cite{burr90}
 nevertheless, it remains as a provocative piece of evidence.

 Another point that should be kept in mind is that $\nenu$ oscillations
 appear \cite{kern95} to be consistent or even favoured by the data when
 the analysis is made model independent.
 This may call for a $\numu$ spectrum of energies below those predicted
 so far but, as has been remarked recently \cite{jank96}, this may be obtained
 if non-elastic \nnn\ scatterings in the transport sphere are taken in
 consideration.

 A typical signature of the assumed $\nu$ mixing
 pattern is the non-observation of the initial $\nu_e$ burst.
 It turns more subtle however when one approaches the non-adiabatic frontier
because the level crossing probability \cite{haxt86,park86} depends
 exponentially on the inverse of the \nnn\ energy.
 Thus, almost-adiabatic transitions for $10\mev$ \nnns\ may be consistent
 with the survival of the highest energy \nnns\ present in the electron
 capture $\nu_e$ burst.
 The survival of $32\mev$ $\nu_e$s was studied in \cite{mina88} and in
 particular, its consistency with the solar non-adiabatic solution.

\section{Conclusions}

 The main result of this paper is that despite the fact that true \gs\
 bosons only have derivative couplings with matter particles, one can
 prove that coherent Goldstone
 fields develop whenever the matter charges they are associated with are
 not conserved in macroscopic scales.
 The equations of motion of the \gs\ bosons
 are equivalent to the conservation law of the No\"ether currents associated
 to nonlinear symmetries of the type $\ffi \rarrow \ffi + \alpha$,
 these currents are a sum of a Goldstone boson piece and a matter current,
 and, whenever the quantum numbers of the matter particles are not conserved,
 the \gs\ fields develop compensating terms.
 Examples are where this may take place in a macroscopic proportion and as
 a result, the matter particles become subject to long range forces.

 In part because the potential energies are inversely proportional to the
 square symmetry breaking scale and on the other hand, proportional to the
 time rates of
 particle reactions, the situations of interest are special, but not unique!
 Here I exercised a scenario where Lepton flavour violation is realized
 through \nnn\ oscillations in a Supernova with large enough intensity 
 to alter the $\nu$ potentials and oscillation patterns, provided that
 the symmetry breaking scale is less than $1\tev$.
 The caracteristic feature of those kind of oscillations is their correlation
 with the periods of largest $\nu$ fluxes thus providing a challenge
 to better understand the time evolution of the \nnn\ signal and the flavour
 dependent energy spectra as well.
 It seems worth to explore other physical systems and symmetries.

\acknowledgements
This work was supported in part by the project ESO/P/PRO/1127/96.



%
%

%
%

\end{document}